\documentclass[12pt,preprint]{aastex}
 \usepackage{apjfonts}
\usepackage{threeparttable}
\usepackage{epstopdf}
\usepackage{longtable}
\usepackage{amsmath}

\usepackage[bookmarks,breaklinks]{hyperref}
   \hypersetup{
     colorlinks,
     citecolor=blue,
     linkcolor=blue,
    }
\begin{document}
\shorttitle{Fine-scale structure of 3C 279}
\shortauthors{Lu et al.}
\title{\uppercase{Fine-scale structure of the quasar 3C 279 Measured with 1.3 \lowercase {mm} very long baseline interferometry}}
\author{Ru-Sen Lu \altaffilmark{1},
          Vincent L.\ Fish\altaffilmark{1},
          Kazunori Akiyama\altaffilmark{2,3}, 
          Sheperd S.\ Doeleman\altaffilmark{1,4},
          Juan C.\ Algaba\altaffilmark{5},
          Geoffrey C.\ Bower\altaffilmark{6},
          Christiaan Brinkerink\altaffilmark{7},
          Richard Chamberlin\altaffilmark{8},
          Geoffrey Crew\altaffilmark{1},
          Roger J. Cappallo\altaffilmark{1},
          Matt Dexter\altaffilmark{6},
         Robert Freund\altaffilmark{9},
         Per Friberg\altaffilmark{10},
         Mark A. Gurwell\altaffilmark{4},
          Paul T. P. Ho\altaffilmark{5},
         Mareki Honma\altaffilmark{2,11},
         Makoto Inoue\altaffilmark{5},
         Svetlana G. Jorstad\altaffilmark{12,13},
          Thomas P.\ Krichbaum\altaffilmark{14},
          Laurent Loinard\altaffilmark{15},
          David MacMahon\altaffilmark{6},
          Daniel P.\ Marrone\altaffilmark{9},
          Alan P. Marscher\altaffilmark{12},
         James M.\ Moran\altaffilmark{4},
         Richard Plambeck\altaffilmark{6},
         Nicolas Pradel\altaffilmark{5},
         Rurik Primiani\altaffilmark{4},
         Remo P.\ J.\ Tilanus\altaffilmark{10,16},
         Michael Titus\altaffilmark{1},
         Jonathan Weintroub\altaffilmark{4},
         Melvyn Wright\altaffilmark{6},
          Ken H.\ Young\altaffilmark{4},
          Lucy M.\ Ziurys\altaffilmark{9}
}
\email{rslu@haystack.mit.edu}
\altaffiltext{1}{Massachusetts Institute of Technology, Haystack
  Observatory, Route 40, Westford, MA 01886, USA}
 \altaffiltext{2}{National Astronomical Observatory of Japan, Osawa 2-21-1, Mitaka, Tokyo 181-8588, Japan}
 \altaffiltext{3}{Department of Astronomy, Graduate School of Science, The University of Tokyo, 7-3-1 Hongo, Bunkyo-ku, Tokyo 113-0033, Japan}
\altaffiltext{4}{Harvard-Smithsonian Center for Astrophysics, 60 Garden
   St., Cambridge, MA 02138, USA}
  \altaffiltext{5}{Institute of Astronomy and Astrophysics, Academia
   Sinica, P.O. Box 23-141, Taipei 10617, Taiwan, R.O.C.}
 \altaffiltext{6}{University of California Berkeley, Dept.\ of
   Astronomy, Radio Astronomy Laboratory, 601 Campbell, Berkeley, CA
   94720-3411, USA}
    \altaffiltext{7}{ Department of Astrophysics, IMAPP, Radboud University Nijmegen, P.O. Box 9010, 6500 GL Nijmegen, The Netherlands}
  \altaffiltext{8}{Caltech Submillimeter Observatory, 111 Nowelo Street, Hilo, HI 96720,
USA}
  \altaffiltext{9}{Arizona Radio Observatory, Steward Observatory,
   University of Arizona, 933 North Cherry Ave., Tucson, AZ 85721-0065,
   USA}
 \altaffiltext{10}{James Clerk Maxwell Telescope, Joint Astronomy Centre,
   660 North A'ohoku Place, University Park, Hilo, HI 96720, USA}
 \altaffiltext{11}{The Graduate University for Advanced Studies, Osawa 2-21-1, Mitaka, Tokyo, 181-8588, Japan}
  \altaffiltext{12}{Institute for Astrophysical Research, Boston University, Boston, MA  02215}
   \altaffiltext{13}{Astronomical Institute, St. Petersburg State University, Universitetskij Pr. 28, Petrodvorets, 198504 St. Petersburg, Russia}
  \altaffiltext{14}{Max-Planck-Institut f\"{u}r Radioastronomie, Auf dem
   H\"{u}gel 69, D-53121 Bonn, Germany }
    \altaffiltext{15}{Centro de Radiostronom\'{\i}a y Astrof\'{\i}sica, Universidad Nacional Aut\'onoma de M\'exico, 58089 Morelia, Michoac\'an, M\'exico}
 \altaffiltext{16}{Netherlands Organization for Scientific Research, Laan
   van Nieuw Oost-Indie 300, NL2509 AC The Hague, The Netherlands}
\begin{abstract}

We report results from 5-day VLBI observations of the well-known quasar 3C 279 at 1.3\,mm (230\,GHz) in 2011. The measured nonzero closure phases on triangles including stations in Arizona, California and Hawaii indicate that the source structure is spatially resolved. We find an unusual inner jet direction at scales of $\sim$1 parsec extending along the northwest-southeast direction (PA = $127^{\circ}\pm3^{\circ}$), as opposed to other (previously) reported measurements on scales of a few parsecs showing inner jet direction extending to the southwest. The 1.3\,mm structure corresponds closely with that observed in the central region of quasi-simultaneous super-resolution VLBA images at 7\,mm. The closure phase changed significantly on the last day when compared with the rest of observations, indicating that the inner jet structure may be variable on daily timescales. The observed new direction of the inner jet shows inconsistency with the prediction of a class of jet precession models. Our observations indicate a brightness temperature of $\sim 8\times10^{10}$ K in the 1.3\,mm core, much lower than that at centimeter wavelengths. Observations with better {\it uv} coverage and sensitivity in the coming years will allow the discrimination between different structure models and will provide direct images of the inner regions of the jet with 20--30 $\mu$as (5--7 light months) resolution.

\end{abstract}

\section{Introduction}

Highly beamed relativistic jets associated with active galactic nuclei (AGN) are believed to be powered by accretion of the central super massive black holes and/or their spin. The creation mechanism of these jets, however, remains an enigma. VLBI observations at short millimeter wavelengths provide the highest angular resolution achievable in astronomy and allow us to explore the compact core regions of these jets, which are self-absorbed at longer wavelengths. In recent years, VLBI observations at 3.5\,mm have been regularly performed either with the Global Millimeter VLBI array (GMVA) or with the standalone Very Long Baseline Array (VLBA). At 1.3\,mm, significant progress has been achieved in the last few years with the Event Horizon Telescope (EHT)~\citetext{ Sgr A*: \citealt{Doeleman2008}, \citealt{2011ApJ...727L..36F},  1921-293: \citealt{2012ApJ...757L..14L}, M87:  \citealt{2012Sci...338..355D}}. 

At a redshift of $z=0.536$~\citep{1996ApJS..104...37M}, the quasar 3C~279 (B1253-055, 1 mas corresponding to 6.31\,pc and 0.1\,mas/yr corresponding to a transverse velocity of 3.2\,c, assuming a standard cosmological model with H$_0$ = 71\,km\,s$^{-1}$ Mpc$^{-1}$, $\Omega_M$ = 0.27, $\Omega_\Lambda$ = 0.73) is one of the brightest radio sources. It is the first source in which superluminal motion was discovered ~\citep{1971Sci...173..225W}, with its jet oriented very close to the line of sight \citep[$0.1^{\circ}$ -- $5.0^{\circ}$,][]{2013AJ....145...12B}. The high brightness and compactness made 3C 279 one of the primary targets for high-frequency VLBI fringe detections \citep{2002A&A...390L..19G,1997A&A...323L..17K}. It is known for its rapid variations across the entire electromagnetic spectrum with various timescales from hours/days to years, and has been monitored intensively at various wavelengths \citep[e.g.,][]{2008A&A...492..389L,2010A&A...522A..66C,2012ApJ...754..114H}.

The large-scale radio structure of 3C~279 is characterized by a compact flat-spectrum core with a jet extending to about 5\arcsec\,along a position angle of $205 ^{\circ}$ (measured north through east) and a radio lobe extending to the northwest about 10\arcsec\,away \citep[e.g.,][]{1983ApJ...273...64D}. \citet{2002ApJ...581L..15C} detected optical and ultraviolet emission from the kiloparsec-scale jet and found a bright knot coincident with a peak in the radio jet $\sim$ 0.6\arcsec\,from the nucleus.

3C 279 has been closely monitored with VLBI over 4 decades, yielding a wealth of detailed information about the jet on parsec scales \citep[e.g.,][]{1971ApJ...170..207C,1979ApJ...229L.115C,1981AJ.....86..371P,1989ApJ...340..117U,2001ApJS..134..181J,2001ApJS..133..297W,2003ApJ...589L...9H,2005AJ....130.1418J,2008ApJ...689...79C}. On VLBI scales, the structure is dominated by a jet extending to the southwest, with multiple projected apparent speeds and position angles. Polarimetric observations have detected both linearly and circularly polarized emission from the parsec-scale jet of 3C 279 \citep[e.g.,][]{1995AJ....110.2479L,1998Natur.395..457W,2000ApJ...533...95T,2001ApJ...553L..31A,2001ApJ...550L.147Z,2009ApJ...696..328H}. The first detection of circular polarization has been used to argue for an electron-positron composition of the jet~\citep{1998Natur.395..457W}. 

In this paper we present results from high-resolution 1.3\,mm VLBI observations of 3C~279. Section \ref{sect:observation}--\ref{sect:reduction} summarizes the EHT observations at 1.3\,mm, data reduction, and calibration. $\S$~\ref{sect:results} describes the results from these observations, and the source structure at 7\,mm from VLBA observations, followed by discussion in $\S$~\ref{sect:discussion}. $\S$~\ref{sect:conclusion} summarizes our conclusions.

\section{Observations}
\label{sect:observation}
An ensemble of sources (3C~273, M~87, 3C~279, 1633$+$382, 3C~345, NRAO~530, Sgr~A*, 1749+096, 1921-293, BL~Lac, 3C~454.3) was observed with the EHT during the nights of 2011 March 29 and 31 and 2011 April 01, 02, and 04 (days 88, 90, 91, 92, and 94). Here we focus on the source 3C~279 and others will be presented elsewhere. Left circular polarization was observed at all sites. Two 480-MHz bandwidths were centered at 229.089 and 229.601~GHz (referred to as the low and high bands, respectively). The DBE1 digital backend designed at MIT Haystack Observatory was used for all single-antenna stations.  Beamformers at the Submillimeter Array (SMA) and Combined Array for Research in Millimeter-wave Astronomy (CARMA) are based on the DBE1 architecture as well.  Both the DBE1 and beamformer systems channelize the data into 15 32-MHz channels. Hydrogen masers were used as timing and frequency references at all sites, with the exception that the 10~MHz reference signal for the 1024~MHz sampler clock in the digital backends at CARMA was erroneously derived from a local rubidium oscillator instead of the hydrogen maser on days 88-92;
the hydrogen maser was used on day 94. Data were recorded onto modules of hard drives using the Mark 5B+ system and correlated with the Haystack Mark 4 VLBI correlator using 32 lags and an accumulation period of 1~s.

The array consisted of seven stations located at three different 
sites: the CARMA site in California, Mauna Kea in Hawaii, and Mount
Graham in Arizona.  The stations are indicated by the one-letter codes
used hereafter (Table~\ref{Table:array}):

\begin{description}

\item[C]  A single 10.4-m CARMA antenna (C4).  The CARMA beamformer
  was used in a ``passthrough'' mode as the backend.

\item[D] A single 10.4-m CARMA antenna (C1) using a DBE1 backend.

\item[F] The phased sum of seven CARMA antennas.  Signals from three
  10.4-m and four 6.1-m antennas were added using the CARMA
  beamformer.  Only the high band was phased.  Station F replaced
  station C in the high band when it was used on days 91-94.

\item[J] The James Clerk Maxwell Telescope (JCMT), using a DBE1
  backend.  Station J was used as a standalone antenna on day 88 only.

\item[O] The Caltech Submillimeter Observatory (CSO), using a DBE1
  backend.  Station O was used as a standalone antenna on days 90--94.

\item[P] The phased sum of seven SMA antennas plus
  either the CSO (day 88) or the JCMT (days 90--94).

\item[S] The Arizona Radio Observatory Submillimeter Telescope (SMT)
  on Mount Graham.

\end{description}

\begin{table*}[ht!]
\centering
\caption{Array description\label{Table:array}.}
\begin{tabular}{lcccc}
\hline
\hline 
Facility&Code&Effective aperture&Days&Note\\
 &&(m)&&\\
\hline
 CARMA&C&10.4&88--94&single dish, high band replaced by F on days 90--94\\
 CARMA&D&10.4&88--94&single dish\\
 CARMA&F&21.8&91--94&$3\times10.4\,{\mathrm m} + 4\times6.1\,{\mathrm m}$ (high band only)\\
 Hawaii&J&15.0&88&JCMT standalone\\
 Hawaii&O&10.4&90--94&CSO standalone\\
 Hawaii&P&19.0/21.8&88--94&SMA ($7\times6.0\,{\mathrm m})$ $+$ CSO (day 88)/JCMT (days 90--94)\\
 SMT&S&10.0&88--94&\\
\hline
\end{tabular}
\end{table*}

\subsection{Weather}

Observations were triggered at all sites several hours before the start of each night's schedule based on expected weather conditions. On Mauna Kea 90\% of zenith opacity measurements over the course of the five nights of observations fell into the range of 0.04--0.15.  The 90\% opacity ranges at the SMT and CARMA were 0.11--0.34 and 0.23--0.48, respectively.  Atmospheric coherence was poor at the CARMA site on day 92, with effects as noted in the following sections.

\section{Data Reduction and Calibration}
\label{sect:reduction}
We reduced the data using the Haystack Observatory Postprocessing System (HOPS). Variable tropospheric conditions at each site
introduce phase errors that cause signal loss for integrations longer than the coherence time of the atmosphere, which typically
ranges from 1~s to about 20~s, depending on the weather \citep{2001AJ....121.2610D,Doeleman2008}.  HOPS tasks are designed to deal with data that are segmented on these short timescales, increasing the signal-to-noise ratio (S/N) of detection relative to
coherently-integrated data and accounting for amplitude losses due to atmospheric decoherence \citep{1995AJ....109.1391R}.

Initial baseline fringe fitting was done without segmentation using the HOPS task \texttt{fourfit}.  Detections with high S/N were used to determine several important quantities for further processing.  First, the phase offsets between the 32-MHz channels within each band were determined. Second, approximate atmospheric coherence times maximizing the S/N of detection were calculated to guide further incoherent fringe searching.  Third, the residual singleband delay, multiband delay, and delay rate were used to set up narrow search windows for each source to assist in fringe finding.

Scans that produced weak detections or no detection at all were reanalyzed with \texttt{fourfit}. Data were then segmented at
4--6~s at a grid of values of multiband delay and delay rate, and the amplitudes were time-averaged at all grid cells. This technique, known as incoherent averaging, allows weaker fringes to be detected in the presence of rapidly variable tropospheric delays.
Finally, detected fringes were segmented at a cadence of 1~s and averaged to produce noise-debiased estimates of the correlation coefficients. Segmented bispectra were also formed and then averaged to construct the closure phase.

The integration time of the incoherent averaging in the final analysis is about 4\,minutes per scan for days 88 and 90, 5\,minutes for days 91 and 92, and 3\,minutes for day 94. All baselines yielded detections on 3C 279 with typical S/Ns of up to 300--400 on the intra-site baselines and 4--6 on the longest baselines (between Arizona and Hawaii). Variations in S/N are mainly due to source structure. However, partially due to known poor atmospheric conditions at some sites, there were some non-detections.

\subsection{Issues Relating to the Mixed Hydrogen Maser/Rubidium Setup}
\label{sect:reduction_timing}
The use of the rubidium standard to drive the sampler clocks at CARMA on days 88-92 introduced small delay and frequency drifts
into the data.  A detailed investigation of affected scans showed that these could be compensated for by introducing an additional
multiband delay drift term along with a proportional local oscillator frequency offset. The value of this delay drift was different from scan to scan and ranged in magnitude up to about 32~ps\,s$^{-1}$.  An analysis of high S/N scans indicated that the delay drift could be characterized as being approximately linear with time over the course of a scan lasting several minutes.  A modified version of \texttt{fourfit} was produced to allow a user-specified delay drift to be removed before fringe fitting.

Since the same timing references were used at all CARMA telescopes, data on CARMA-CARMA baselines (CD and FD) were unaffected. Additionally, a single delay drift value per scan is sufficient to characterize the effects of this setup for both bands and all
baselines including exactly one CARMA station. Determination of delay drift values was done on a scan-by-scan basis by maximizing the S/N of the SMT-CARMA baselines (SD and either SC or SF), which are generally the highest S/N baselines to CARMA due to the fact that the relatively short baseline length does not resolve out as much source emission as the Hawaii-CARMA baselines do. 

To the extent that the delay drifts were not linear over the course of a scan, it is possible that the calibrated VLBI amplitudes on the CARMA baselines slightly underestimate the true values. An analysis of the variation of sub-scan delay solutions showed a typical rms scatter of about 0.2\,ns during a 4 min scan. The scatter of the delay corresponds to a de-correlation of $\sim$ 2\,\%, caused by a phase change of about $35^{\circ}$ across the 480\,Hz band, which is less than the uncertainties imposed by calibration.

\begin{figure*}[ht!]
\epsscale{0.60}
\plotone{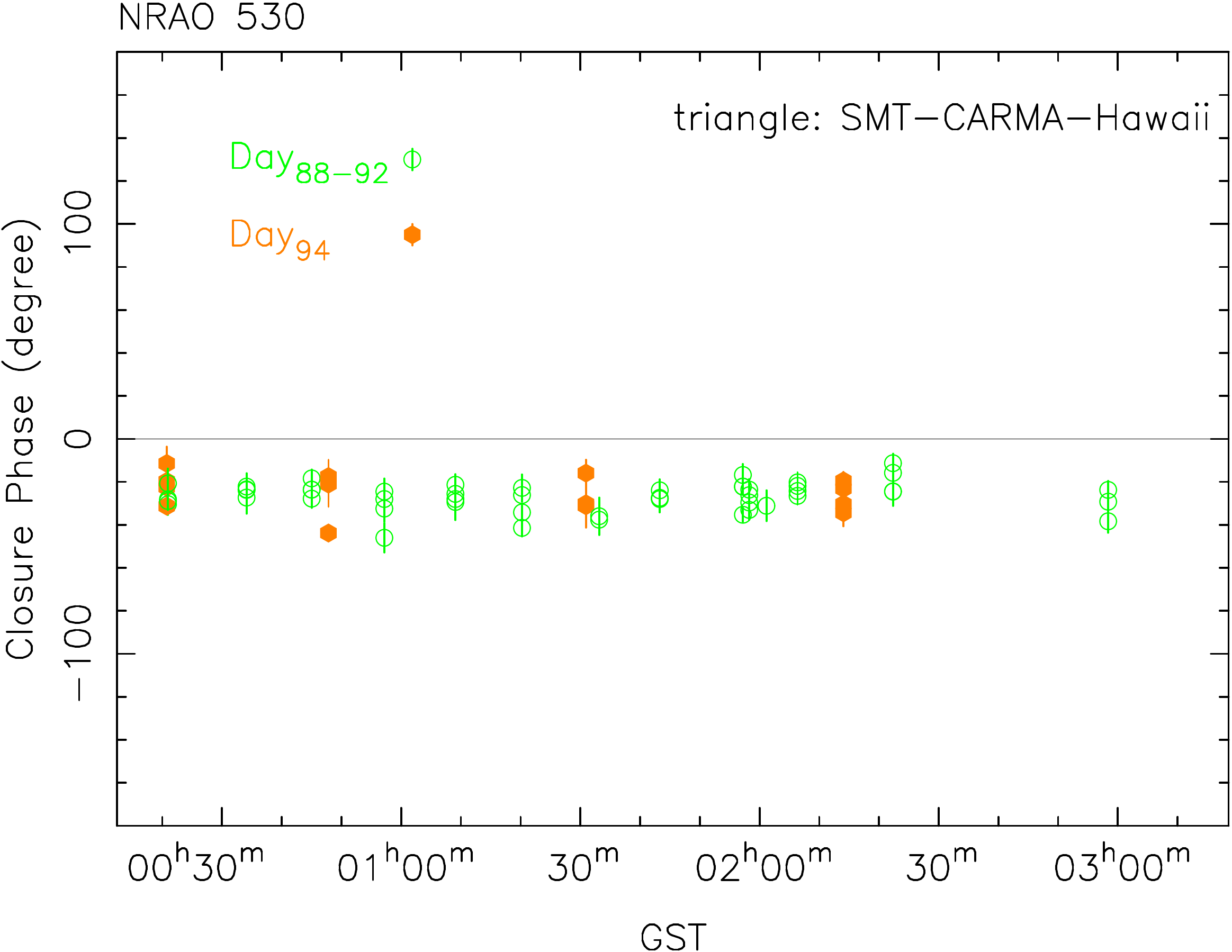}
\caption{Closure phases of NRAO 530 on the SMT-CARMA-Hawaii triangle.}
\label{Fig:cphase_NRAO530}
\end{figure*}

We verified that closure phases on triangles including a CARMA station are unaffected, since station-based clock errors do not typically introduce non-closing phase errors. The corrections applied in the modified version of \texttt{fourfit} are likewise station-based, so phase rotations applied on one baseline including a CARMA station are exactly compensated by a negative rotation on the other baseline involving CARMA in a triangle. Change of the delay drift around the best-fit values does not introduce significant changes in closure phase. Analysis of sub-scans at shorter intervals with equal duration also show consistency of closure phases with scan-averaged values, indicating that a single delay drift per scan is sufficient. As an example, Figure~\ref{Fig:cphase_NRAO530} compares the closure phase of NRAO~530 on days 88-92 with that on day 94 on the SMT-CARMA-Hawaii triangle; Consistent closure phases were measured on the five days of the observation.

\subsection{A Priori Calibration}

VLBI correlation coefficients are converted to flux densities by multiplication by the geometric mean of the opacity-corrected system
equivalent flux density (SEFD) of the two telescopes on a baseline. The SEFD in turn depends on both the system temperature
($T_\mathrm{sys}$) of each scan and the gain of the telescope (or, equivalently, the geometric area and the aperture efficiency).

The aperture efficiencies of the JCMT and the SMT were computed from planet observations in 2009 using a setup identical to the one used for 2011 \citep{2011ApJ...727L..36F}. Planet observations were also used to determine the CSO aperture efficiency. System temperatures were obtained using a vane calibration technique. System temperature measurements were taken before each VLBI scan at all sites except at the CSO, where fewer measurements were taken.  Comparison of the measured $T_\mathrm{sys}$ values at the CSO with those of the nearby JCMT showed a strong correlation, and an empirical fit was used to
estimate the system temperature of the CSO when values were missing. System temperatures at CARMA were measured using a chopper wheel before each VLBI scan, then updated continuously during the scan under the assumption that the receiver gain is constant.

Phase switching and lobe rotation were disabled for the single CARMA antennas, preventing simultaneous interferometric and VLBI
observations, so SEFDs for stations C and D were determined from CARMA correlator data obtained immediately before each VLBI scan. Self-calibration was used to determine gain corrections for each scan, which were then used to produce an opacity-corrected SEFD.  For station F, complex voltage correction factors were calculated from self-calibration every 10~s and used in constructing the phased sum. As a result, the SEFDs for station F implicitly take into account losses due to imperfect phasing.  The median phasing efficiency exceeded 70\% on days 91 and 94 but was only 50\% on day 92 due to increased atmospheric turbulence.

On Mauna Kea the phased array system installed at the SMA aggregated the collecting area of seven SMA antennas with the CSO as the eighth antenna on day 88 and the JCMT as the eighth on the last four days of the observation. The phased array system estimates correcting phases on a scan by scan basis using a seven baseline ``calibration correlator''. The phasing efficiency for each calibration correlator frequency is estimated as the absolute value of the vector sum divided by the scalar sum of the calibration correlator visibilities. The sums are taken over the calibration correlator baselines to the reference antenna.  These estimated phasing efficiencies for each frequency bin are then averaged over frequency to produce a single phasing efficiency estimate for each calibration correlator scan.  An SEFD for the corresponding phased sum ($SEFD_{phased}$) is computed, by first computing the SEFD for each dish ($SEFD_{i}$) in the phased array as $SEFD_{i}=2\times T_{sys_{i}} \times G_{D_{i}}$, where $T_{sys_{i}}$ is the measured DSB system temperature and $G_{D_{i}}$ is the dish gain (in Janskys per Kelvin) for antenna $i$.
We use measured gain values of 130 Jy/K for each SMA dish, 29.5 Jy/ K for the JCMT, and 62.2 Jy/K for the CSO. The SEFD for the phased sum is then computed as $SEFD_{phased}=(\eta\sum\limits_{i}\frac{1}{SEFD_{i}})^{-1}$, where $\eta$ is the estimated phasing efficiency.

\subsection{Flux Density Measurements}

In the gaps between VLBI scans, the SMA and CARMA also obtained flux density measurements of target sources.  SMA flux densities are available for a subset of sources on days 88, 91, and 92. 
Titan and MWC\,349a were the primary calibrators on days 88 and 92, and constant flux density was assumed for 3C\,273 and 3C\,274 for primary flux density calibration on day 91~\citep[see][for more details]{2007ASPC..375..234G}. CARMA flux densities are available for most sources on days 90 and 91\footnote{http://mmarray.org/memos/carma\_memo61.pdf}. Uranus was the primary flux calibrator at CARMA.  In Mar/Apr 2011 Uranus was a 3.34$\times$3.27 arcsecond disk, assumed to have a uniform brightness temperature of 102 K.  Antenna amplitude gains derived from a self-calibration solution to the Uranus source model were applied to other sources in order to estimate their fluxes. Uranus is much smaller than the primary beamwidths of the CARMA antennas at 229 GHz, which are 30\arcsec and 50\arcsec  for the 10-m and 6-m telescopes, respectively. The flux densities derived at CARMA were somewhat higher than those derived at the SMA, by a median value of 8\,\%. This may be because Uranus was observed at the end of the VLBI schedule, after sunrise, when the antenna focus and pointing degrade slightly, or may be due to the uncertainties in the absolute flux scales.
 
Most of the EHT target sources are pointlike when observed at the resolution of the SMA or CARMA.  Each scan with the EHT array includes two high S/N intra-site baselines (JP or OP at Mauna Kea and CD or FD at CARMA).  Correlation coefficients measured on the CARMA-CARMA VLBI baseline appear to be a few percent higher than the corresponding correlation coefficient obtained by the CARMA correlator, possibly indicating that the VLBI processing, done with an accumulation period of 1~s, is somewhat more immune to atmospheric coherence losses than the CARMA correlation, which uses an averaging time of 10~s. This is especially true on day 92, when the unstable atmosphere caused the CARMA correlation amplitudes to have a large scatter relative to the VLBI-derived quantities and the CARMA phased array to have correlation coefficients that are biased very low with respect to the VLBI processing.

We have assumed the ``zero-spacing'' flux densities measured on the intra-site baselines equal to the average of the CARMA interferometer measurements on days 90 and 91.

\subsection{Other Known Issues}

Two additional instrumental effects are noted with the phased array at the SMA.  The a priori amplitudes are corrected for these effects before further processing.

The high-band amplitudes measured on baselines including station P are systemically lower than the low-band amplitudes. The magnitude of this effect is comparable on all baselines to station P, and it is not seen on any other baseline.  The source of the high-band loss has not been identified, but could be caused by a gain slope across the band.  We have corrected empirically for the loss by multiplying the a priori high-band amplitudes on baselines including station P by the ratio of the low-band and high-band flux densities on the JP baseline (day 88) or OP baseline (days 90--94).  Detections on these intra-site baselines have very high S/N, and therefore we do not expect this correction to introduce additional errors into the the high-band amplitudes.  In the rare instance that neither the JP nor OP baseline is available (for instance, if the other antenna on Mauna Kea missed a scan), the SP baseline is used instead. The average amount of this correction is about 34\,\%.

Even after correcting the high-band amplitudes, the amplitudes on the XP baseline in both bands are lower than on the XJ or XO baseline, where X represents any other station.  The a priori amplitudes on the XP and XJ/XO baselines are tightly correlated, but the XP amplitudes must be increased by an average of 23\% for the amplitudes to match. This effect appears to be related to station P rather than either station J or O, as the amplitudes obtained on baselines to station P are lower than the corresponding amplitudes to both station J (on day 88) and station O (on days 90--94). The source of this loss factor is not well understood, but possible reasons may include inaccuracies in calculating the antenna gains and phasing efficiency. We have removed its effect by multiplying all XP amplitudes by 1.23.

\subsection{Gain Correction}

Calibration is completed by using amplitude self-calibration.  For each scan and band, gains are calculated for each station to produce maximal consistency of the data with each other and with the ``zero-spacing'' flux densities measured with the SMA and CARMA. The method is similar to that described in \citet{2011ApJ...727L..36F}, except that all baselines are used and weighted by the data S/N in computing the gain correction coefficients. After gain correction, the calibrated intra-site VLBI amplitudes (e.g., CD) are equal to the flux density measured by CARMA, and amplitudes of substantively identical VLBI quantities (e.g., SC and SD, low and high bands) are made as equal as possible.

\section{Results}
\label{sect:results}
\subsection{Source Structure Modeling at 1.3 mm}
\begin{figure*}[ht!]
\epsscale{1.0}
\plotone{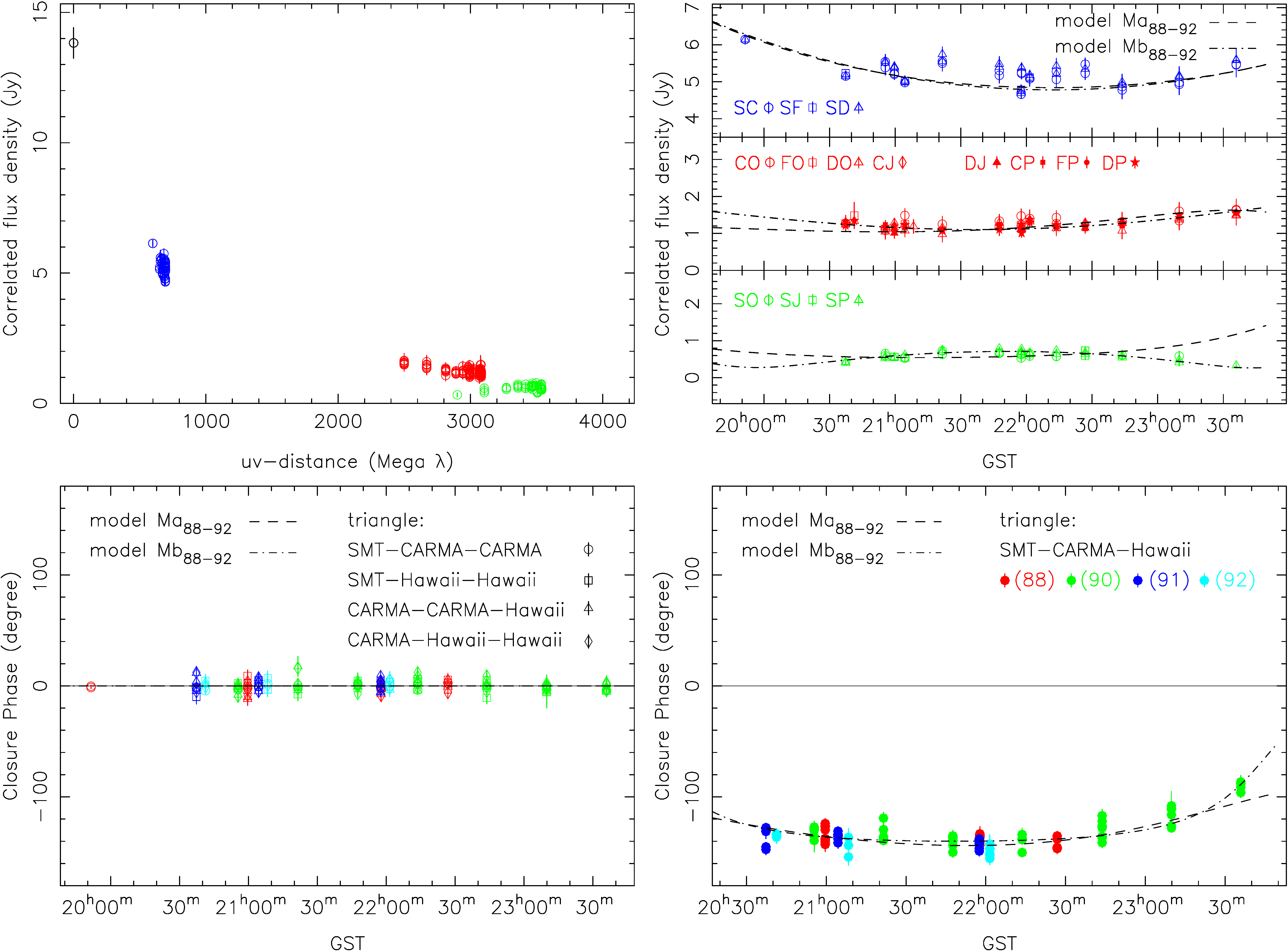}
\caption{Measured correlated flux density and closure phase for days 88--92. Upper panel (left): correlated flux density as a function of {\it uv} distance. Since the source structure is not circularly symmetric, the correlated flux density will change for a given {\it uv} distance as the Earth rotates. Upper panel (right): correlated flux density as a function of time with comparison of the two models shown in Table~\ref{Table:model}. The correlated flux densities are color-coded by baseline. Lower panels: closure phase as function of time for the trivial triangles (left), which include an intra-site baseline, and non-trivial triangles (right). Detections are color-coded by observing epoch, as indicated in the legend in the lower right panel. The predicted closure phases of the two models (Table~\ref{Table:model}) are also shown.}
\label{Fig:d88}
\end{figure*}

\begin{figure*}[ht!]
\epsscale{1.0}
\plotone{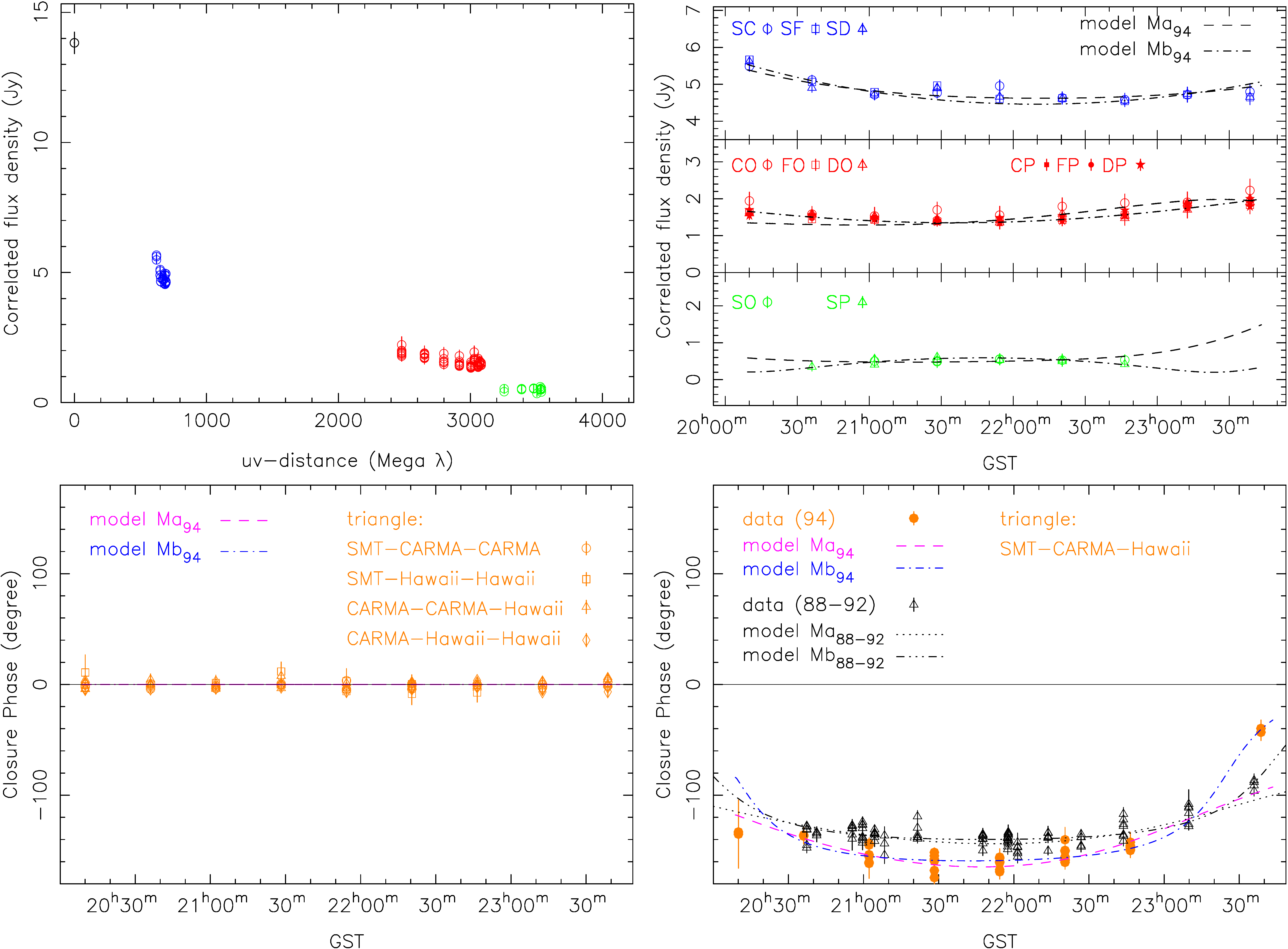}
\caption{Same as Figure~\ref{Fig:d88}, but for day 94. Note that the closure phase data and models for the SMT-CARMA-Hawaii triangle on days 88--92 are overplotted for comparison (lower right, open symbols).}
\label{Fig:d94}
\end{figure*}

The calibrated amplitudes and closure phases (Tables~\ref{tab:flux} and \ref{tab:cphase}) are shown in Figures~\ref{Fig:d88} and \ref{Fig:d94}, respectively. Trivial triangles, which involve an intra-site baseline (JP or OP at Mauna Kea and CD or FD at CARMA), have essentially zero closure phase, indicating a point-like structure at the resolution provided by the intra-site baselines (a few arc seconds).

\begin{table*}
\centering
\caption[Gain-corrected Visibility Amplitudes of 3C 279]{Gain-corrected Visibility Amplitudes of 3C 279}
\label{tab:flux} 
\begin{tabular}{ccccccccc}
\hline
\hline
Day & hh &mm & Baseline & {\it u}& {\it v} & Flux Density & $\sigma$ & Band\\
       &      &       &                & (M$\lambda$)&(M$\lambda$)&(Jy)&(Jy)&\\
\hline
88 & 07 & 26 & CD &     -0.006&     -0.073 &  13.83 &   0.18&high\\  
   & 07 & 26 & SC &   -526.942&    283.963 &   6.14 &   0.12&high\\
   & 07 & 26 & SD &   -526.948&    283.889 &   6.14 &   0.11&high\\
   & 08 & 34 & CD &      0.011&     -0.074 &  13.83 &   0.20&high\\
   & 08 & 34 & CJ &  -2794.721&  -1298.854 &   1.17 &   0.15&high\\
   & 08 & 34 & CP &  -2794.731&  -1298.751 &   1.12 &   0.14&high\\
   & 08 & 34 & DJ &  -2794.732&  -1298.780 &   1.01 &   0.14&high\\
   & 08 & 34 & DP &  -2794.742&  -1298.678 &   1.06 &   0.12&high\\
   & 08 & 34 & JP &     -0.010&      0.103 &  13.83 &   0.06&high\\
   & 08 & 34 & SC &   -598.012&    300.969 &   5.20 &   0.14&high\\
\hline
\end{tabular}
\begin{tablenotes}
\item A portion of this table is shown here to demonstrate its form and content. An online machine-readable version of the full table is available. 
\end{tablenotes}
\end{table*}%

\begin{table}
\centering
\caption[Closure phases of 3C 279]{Closure phases of 3C 279}
\label{tab:cphase}
\begin{tabular}{ccccccc}
\hline
\hline
Day & hh & mm& Triangle &Closure Phase &$\sigma$ &Band \\
&&&&(Degree)&(Degree)&\\
\hline
88 & 08 & 34 & CDP &   -11.9 &   5.8 &  high\\
   & 09 & 32 & CDP &    -1.8 &   4.1 &  high\\
   & 08 & 34 & CDJ &    -0.3 &   5.6 &  high\\
   & 09 & 32 & CDJ &    -2.5 &   6.0 &  high\\
   & 10 & 01 & CDJ &     1.2 &   4.0 &  high\\
   & 08 & 34 & CJP &    -3.2 &   7.2 &  high\\
   & 09 & 32 & CJP &     2.7 &   4.7 &  high\\
   & 08 & 34 & DJP &     3.7 &   4.5 &  high\\
   & 09 & 32 & DJP &     1.1 &   4.1 &  high\\
   & 10 & 01 & DJP &     4.9 &   3.5 &  high\\
\hline
\end{tabular}
\begin{tablenotes}
\item A portion of this table is shown here to demonstrate its form and content. An online machine-readable version of the full table is available. 
\end{tablenotes}
\end{table}%

The source structure at 1.3\,mm was quantitatively parameterized by considering a class of simple circular Gaussian models that fit the calibrated amplitudes and closure phase jointly following \citet{2012ApJ...757L..14L}.  Uncertainties are reported based on the size of the region around the best-fit point in parameter space corresponding to 68.3\,\% probability and the true uncertainties may be somewhat larger. Model fitting was performed for the data on days 88--92 and day 94 separately, because the closure phase on the non-trivial triangles (SMT-CARMA-Hawaii) on day 94 is significantly different than on the other four days (Figures~\ref{Fig:d88} and \ref{Fig:d94}, lower right). Closure phases on other triangles (e.g., CARMA-Hawaii-Hawaii), and for all the other sources on all triangles do not show any sign of change in a similar manner and we do not believe that this change is due to calibration uncertainties ($\S$ \ref{sect:reduction_timing}).  

We started the model fitting with a two-component model since a single-component model is completely ruled out by the
measured nonzero closure phase on the non-trivial triangles. A two-component model (model Ma) yielded a good fit to the amplitudes and nonzero closure phases, representing the most basic structure of the emission. The structure is characterized by a compact, but relatively weak component (A0), and a stronger component (A1) $\sim$130--140\,$\mu$as away at a PA of about -50$^{\circ}$ (Table~\ref{Table:model}). The measured closure phases remove the $180^{\circ}$ of degeneracy in the jet position angle. The two models of Ma$_{88-92}$ and Ma$_{94}$ are very similar to each other, and both indicate a jet direction (regardless of core identification) very different from the typical centimeter jet orientation (i.e., extending to the southwest).

Although the two-component models adequately fit the data, it is interesting to know to what extent a slightly more complicated, three-component model can be constrained. This is encouraged by a few indications that the source structure may be more complicated than assumed, for instance, the two-component model seems to slightly over-predict the flux density on SMT-Hawaii baselines around $23^{\rm h}00$ (Figure~\ref{Fig:d88}, upper right panel) and underestimates the non-trivial closure phase at the end of observations on day 94 (Figure~\ref{Fig:d94}, lower right). For the three-component model (Table~\ref{Table:model}), a few parameters are not well constrained, e.g., the position of the component Mb(2), for both data sets. We therefore introduced an a priori constraint for the position angle of this component (-120$^\circ$, relative to Mb(0)) according to models of the 7\,mm VLBI images (Section~\ref{Sec:7mm}). We also noticed that there are regions in the parameter space that provide alternative three-component models. These three-component models, however, are indistinguishable from the pure perspective of $\chi^2$, indicating that they are only at a marginal level of detail that the current data can support.

\begin{table*}[ht!]
  \centering
  \begin{threeparttable}[b]
  \caption{Model-fit parameters for 1.3\,mm VLBI observations of 3C 279.}
  \label{Table:model}
  \begin{tabular}{lccccccc}
  \hline
  \hline
Model&$\chi^2_{\nu}$ &ID &Flux Density &Radius &PA&Size (FWHM) &T$_{b}$\\
         &                        &     &(Jy)               &($\mu$as)&(degree) &$\mu$as)&(K)\\
\hline
  &&\multicolumn{4}{c}{two-component model (model A)}&& \\ 
\hline
Ma$_{\rm 88-92}$&2.0&0&$3.8\pm0.3$&0&0 &$42\pm3$&$7.7 \times10^{10}$\\
                            &     &1&$10.0\pm0.4$ & $133\pm6$ &  $-54\pm5$  &$73\pm3$&$6.7\times10^{10}$\\
\hline
Ma$_{\rm 94}$&1.8&0&$4.0\pm0.4$&0&0 &$42\pm2$&$8.1\times10^{10}$\\
                       & &1& $9.8\pm0.3$ & $145\pm7$  & $-52\pm4$& $64\pm3$ &$8.5\times10^{10}$\\
 \hline
 &&\multicolumn{4}{c}{three-component model (model B)}&& \\ 
 \hline
 Mb$_{88-92}$&1.4&0& $1.2\pm0.5$&    0 &0    &$13\pm9$   & $2.5\times10^{11}$ \\ 
                         &      &1&$1.0\pm0.4$&$110\pm20$ &$-35\pm3$ &$17\pm$11  &$1.2\times10^{11}$\\ 
                         &      &2&$11.6\pm1.0$&$79\pm29$ &-120\tnote{a}&$168\pm5$&$1.5\times10^{10}$\\ 
 \hline
 Mb$_{94}$& 1.0&0&$1.4\pm0.4$&   0&      0& $11\pm11$  &$4.1\times10^{11}$\\
                     &&1&$1.2\pm0.4$&$91\pm24$& $ -37\pm6$& $13\pm11$ &$2.6\times10^{11}$\\  
                     &&2&$11.2\pm0.7$&$94\pm34$& -120\tnote{a}& $184\pm9$ &$1.2\times10^{10}$\\
\hline
\end{tabular}
\begin{tablenotes}
\item[a] fixed during model fitting
\end{tablenotes}
\end{threeparttable}
\end{table*}

\subsection{Source Structure at 7 mm}
\label{Sec:7mm}
\begin{figure*}[ht!]
\epsscale{1.0}
\plotone{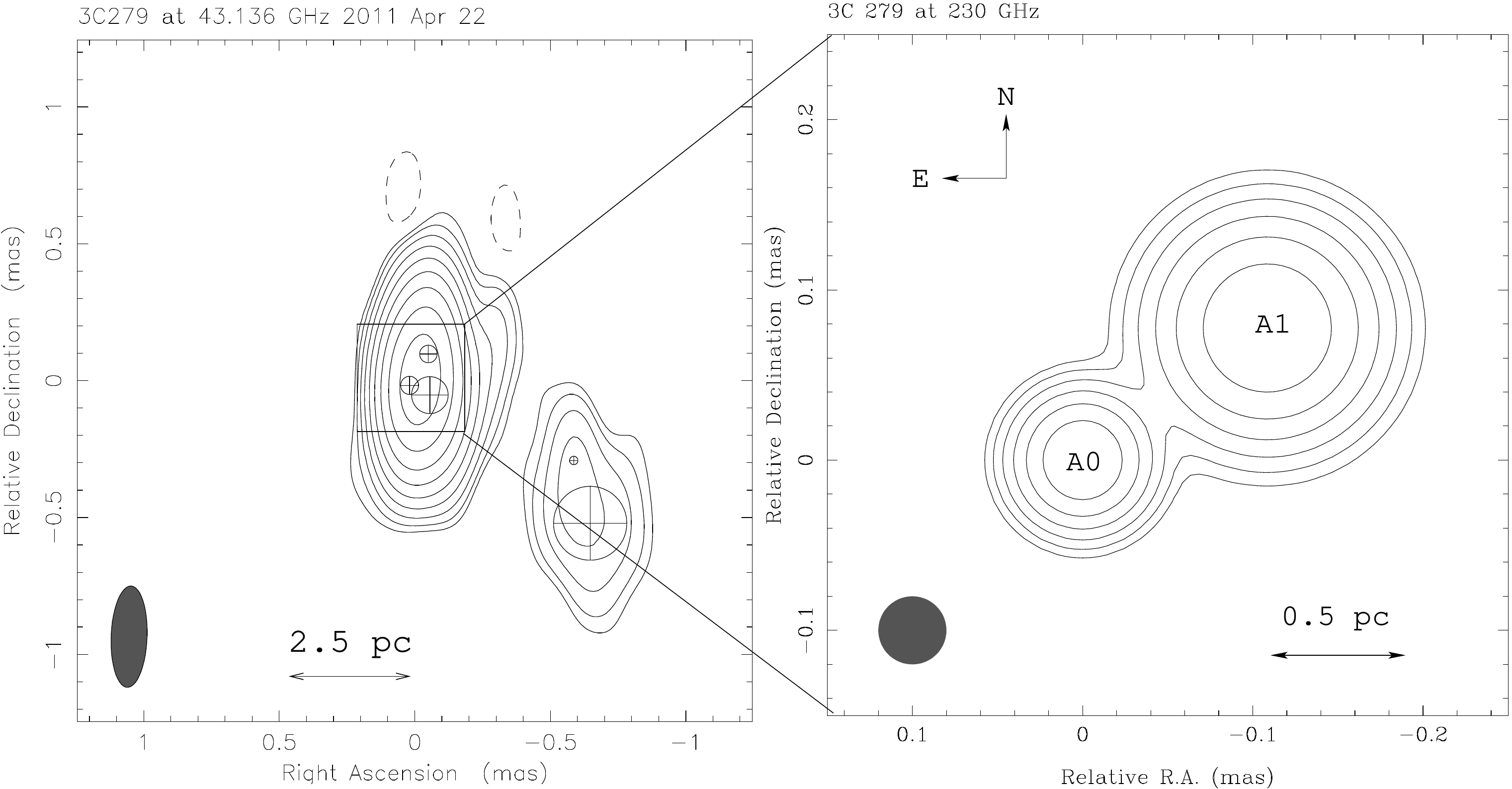}
\caption{Left: Total intensity image of 3C 279 at 7\,mm obtained on 2011 April 22 with model-fit components superimposed~\citep{Jorstad13}. The peak flux density is 15.6 Jy beam$^{-1}$ and contours are drawn at -0.25\,\%, 0.25\,\%, 0.5\,\%, 1\,\%, 2\,\%, 4\,\%, 8\,\%, 26\,\%, 32\,\%, and 64\,\% of the peak. The beam size is approximately $0.37\times0.13$ mas at a position angle of $-2^{\circ}$.  Right: model image of 3C 279 at 1.3\,mm based on data from days 88-92 (model Ma$_{88-92}$). The image is convolved with a circular beam with an FWHM of $20\mu$as. Contours are drawn beginning at 1.6\,\% of the peak brightness of 0.7 Jy/beam and increase by a factor of 2.}
\label{Fig:map}
\end{figure*}

At 7\,mm, the Boston University group monitored 3C 279 with the VLBA roughly monthly. These data have been reduced in the same manner as described in \citet{2005AJ....130.1418J} and will be presented in \citet{Jorstad13}. We turned to these data to check the consistency of the modeled structure at 1.3\,mm after initial model fitting. The images at epochs that bracket the EHT observations begin to show some asymmetry in the core region, evident by the small, but nonzero closure phases on the triangles involving the longest VLBA baselines ($\sim 0.1$ mas resolution). The 2011 April 22 observations at 7\,mm are at the epoch closest to the 1.3\,mm observations (separated by 22 days). Figure~\ref{Fig:map} (left) shows the 7\,mm VLBA image at this epoch, which reveals a triple structure at the central region. The properties of the two compact components (relative position and sizes) at 7\,mm in the central region show remarkable similarity to those of component A0 and A1 (Figure~\ref{Fig:map}, right) with flux densities of 8.8 and 6.1\,Jy. For a three-component model at 1.3\,mm, it is reasonable to assume that B2 corresponds to the extended component at 7\,mm and have similar position given their similar size. Once the a priori PA constraint (similar PA as at 7\,mm) is introduced for component B2, the separation between B2 and B0 shows good consistency with the 7\,mm results.

\begin{figure}[ht!]
\begin{center}
\epsscale{0.50}
\plotone{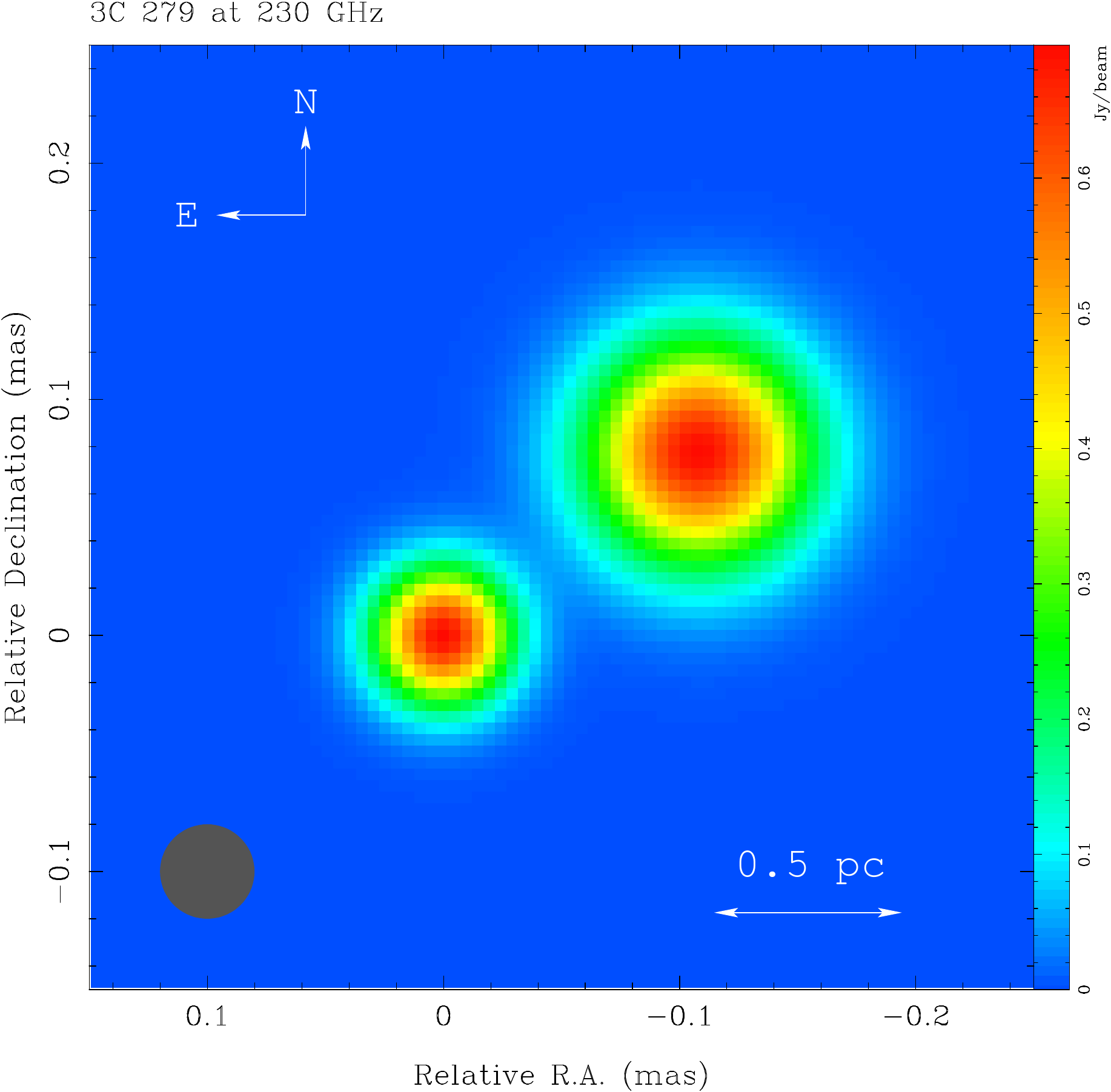}
\caption{The inner jet structure of 3C 279 at 1.3\,mm based on data from days 88-92 (model Ma$_{88-92}$).  The image is convolved with a circular beam with an FWHM of $20\mu$as and the vertical color bar on the right has units Jy/beam.}
\label{Fig:colormap}
\end{center}
\end{figure}

\section{Discussion}
\label{sect:discussion}
The present 1.3\,mm VLBI observations reveal a jet structure of 3C 279 (Figure~\ref{Fig:colormap}) very different from the typical jet direction (i.e., towards south-west) at centimeter wavelengths \citep[e.g.,][]{2005AJ....130.1418J}. The component A1 is likely to be the core, a compact feature located at the upstream end of jets as seen in VLBI images. The nonsimultaneous two-frequency spectral indices $\alpha$ ($S\propto\nu^{\alpha}$) between 43 and 230\,GHz are 0.3 and -0.5 for A1 and A0, suggesting A1 is the core with a flat spectrum, while A0 is optically thin. Determination of the spectral index using data at two frequencies is not a robust estimate hence future high resolution multi-frequency observations are needed to measure the spectral properties and further confirm this scenario. This identification gives an inner jet PA (i.e., the PA of A0 with respect to the core A1) of $\simeq 127^{\circ}\pm3^{\circ}$.

The repeat of observations on successive days in the same GST range permits us to investigate the possible daily variations of flux density and structure of 3C 279. For other blazars, like NRAO 530, the inner jet components were found to be more variable than those further out on daily timescales \citep{2011MNRAS.418.2260L}. The change of closure phases on day 94 provides an opportunity to study the interday variability of 3C 279 on sub-pc scales, which can be caused, e.g.,  by a component brightening.  
From Table~\ref{Table:model}, it seems that the closure phase change can be explained via a small change of source structure and 
variations in how the total flux is divided up, but it cannot be unambiguously attributed to a single model parameter.
Observations at 7\,mm clearly show that A0 is swinging in position angle and moving southwest relative to A1 \citep{Jorstad13}.
In view of these fast changes, more frequent monitoring is needed in the future to follow up these variations. 

Jet components in 3C~279 were known to have notably different speeds and position angles~\citep[e.g.,][]{2001ApJS..133..297W,2004AJ....127.3115J,2008ApJ...689...79C}. This has been interpreted as a consequence of a precessing jet by \citet{1998ApJ...496..172A}. Recently, \citet{2011RAA....11...43Q,2012RAA....12...46Q} proposed a jet-nozzle precession model for 3C 279, in which the jet components move along a common (collimated) curved trajectory precessing with a period of $\sim$ 25 yrs. They predicted the jet ejection angle approaching $\sim -90^{\circ} \pm10^{\circ}$ in 2010-2012. The inner jet position angle seen with 1.3\,mm VLBI is beyond the range of the PA that these models predict (between $\sim -155^{\circ}$ and  $\sim -80^{\circ}$). It should be pointed out that estimating the ejection position angle and time of component B0 is not a straightforward task due to its swing. More erratic changes in apparent speed and direction may naturally result from a small change in the viewing angle \citep{2003ApJ...589L...9H,2004AJ....127.3115J}. The new inner jet direction seen with 1.3\,mm VLBI is probably a similar event, suggesting that the jet points very close to our line of sight. However, the origin of these changes of the jet is still an open question.

Past centimeter VLBI observations have shown that the core of 3C 279 has brightness temperature $> 10^{12} $ K \citep[e.g.,][and references therein]{2000ApJ...537...91P,2001ApJS..133..297W}. Recently, \citet{2008AJ....136..159L} suggested a systematic decrease of jet brightness temperature in compact radio sources towards the mm-VLBI core at 3.5\,mm. In the quasar 1921-293, the low brightness temperature of the 1.3\,mm core is consistent with this trend~\citep{2012ApJ...757L..14L}. For 3C 279, the measured brightness temperature towards the VLBI core at 1.3\,mm is consistent with a lower value of $\sim 8\times10^{10}$ K. 

We expect future observations with improved {\it uv} coverage and sensitivity to provide stricter constraints on the model. Introducing a priori knowledge is able to help models converge, but solutions are non-unique. Moreover, it is not straightforward to do so for current observations due to the fast structure change on the probed spatial scales (sub-parsec for 3C 279) and the lack of observations that are close in time and with matched angular resolution. 
This situation, however, is expected to change soon.

With the planned improvement of millimeter VLBI capability over the next few years, the EHT will allow direct imaging of these inner regions close to the jet origin and will be able to discriminate between these models. Here we consider two synthetic data sets using models Ma$_{\rm 88-92}$ and Mb$_{\rm 88-92}$ as input for future observations of 3C~279 with seven potential stations~\citep{2009ApJ...695...59D,2009ApJ...692L..14F}:
Hawaii, consisting of one or more of the JCMT and SMA phased together into a single aperture; the SMT; CARMA; the Large Millimeter Telescope (LMT) on Sierra Negra, Mexico; the phased Atacama Large Millimeter/submillimeter Array (ALMA); the Institut de Radioastronomie Millim\'{e}trique (IRAM) 30\,m telescope on Pico Veleta (PV), Spain; and the IRAM Plateau de Bure (PdB) Interferometer, phased together as a single aperture. Assumed telescope sensitivities are updated from \citet{2009ApJ...695...59D}. For our simulations, we took phased ALMA as the site in Chile. We note that early imaging studies at 1.3\,mm will likely involve the APEX 12\,m dish, which would sample similar spatial frequencies to the phased ALMA.

\begin{figure}[ht!]
\epsscale{0.5}
\plotone{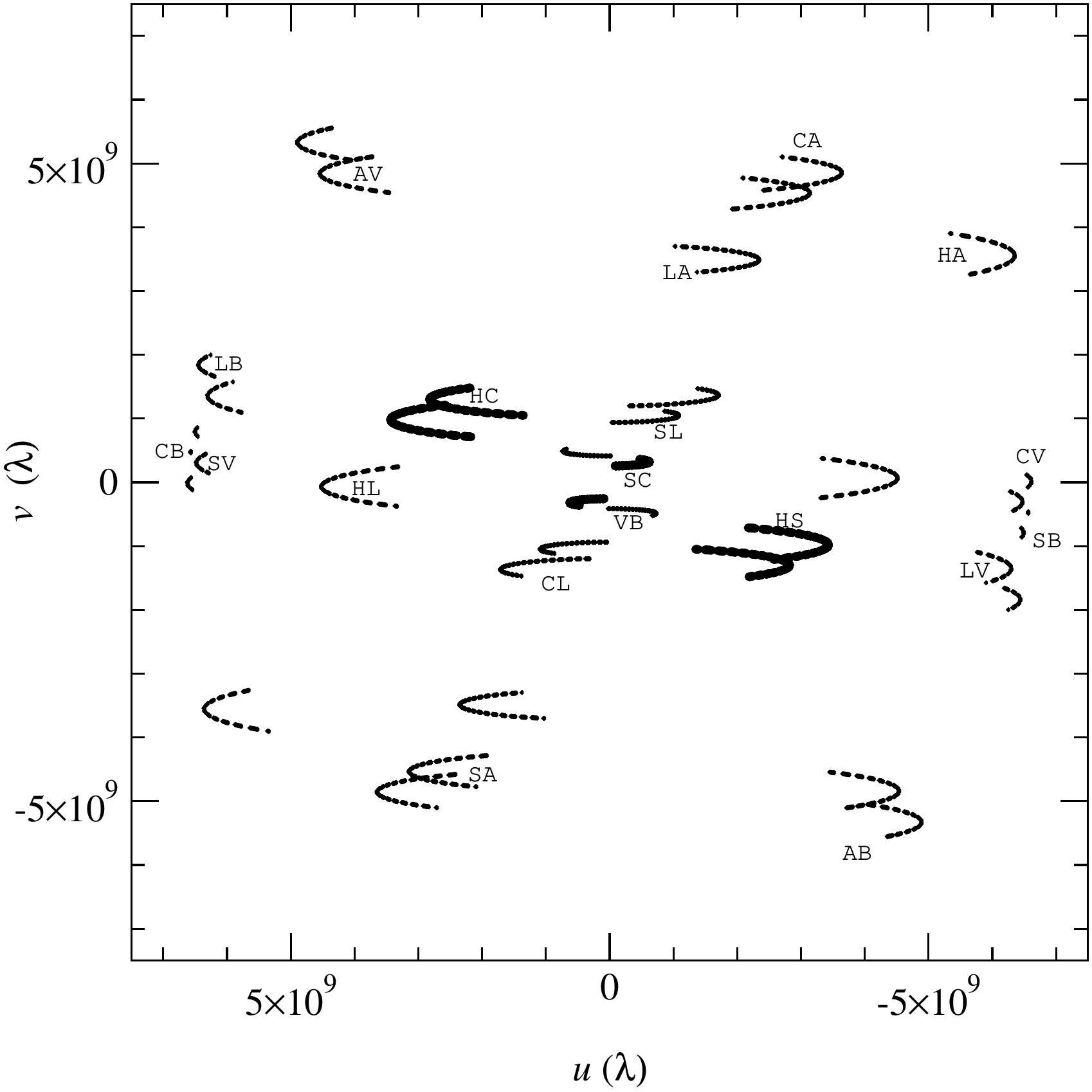}
\caption{{\it uv} coverage for the array used for the simulation. Tracks are labeled by
baseline (H: Hawaii; S: SMT; C: CARMA; L: LMT; A: ALMA; V: Pico Veleta; B: Plateau de Bure Interferometer). The visibility at ({\it -u, -v}) is the complex conjugate of that at (u, v). {\it uv} tracks for the baselines in the 2011 observations are highlighted.}
\label{Fig:2015uv}
\end{figure}

\begin{figure*}[ht!]
\epsscale{1.0}
\plotone{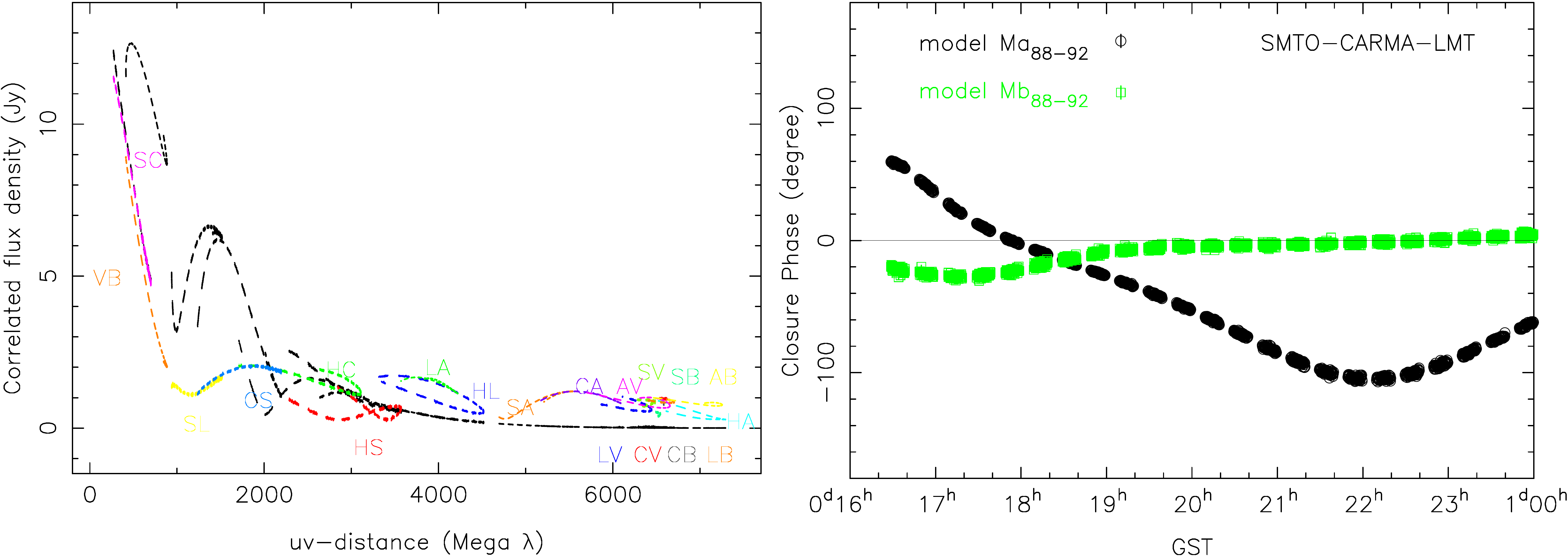}
\caption{Left: correlated flux density as a function of {\it uv} distance for the two models of days 88-92 (Ma$_{88-92}$ and Mb$_{88-92}$, Table~\ref{Table:model}). Points for the model Mb$_{88-92}$ are color-coded and labeled by baseline (same as in Figure~\ref{Fig:2015uv}). Right: plot of the predicted closure phase for the shown time range on the SMT-CARMA-LMT triangle. Simulated data were coherently averaged into 10-s bins.}
\label{Fig:2015ac}
\end{figure*}

The array used in this simulation provides very good {\it uv} coverage on 3C 279 (Figure~\ref{Fig:2015uv}), indicating sufficient imaging capabilities in the near future. We show in Figure~\ref{Fig:2015ac} (left) the predicted visibility amplitudes and the expected closure phases on the SMT-CARMA-LMT triangle as an example (Figure~\ref{Fig:2015ac}, right) for the two models. These models predict different behavior for most of the baselines and triangles, and they can be straightforwardly discriminated. Since these models may be approximating a continuous expanding jet, future observation will enable the direct imaging of the jet structure with improved sensitivity and {\it uv} coverage in the coming years.

\section{Conclusions}
\label{sect:conclusion}
We have presented the first high-resolution VLBI observations of 3C 279 at 230\,GHz.
The sub-parsec-scale emission is dominated by a compact double structure along the northwest-southeast direction. We find radio structures at 1.3\,mm similar to those that were seen in the central core regions of 7\,mm 
super-resolution images. The change of closure phases on the last day during our observations provides opportunities to study sub-parsec-scale jet structure change on daily timescales. If further confirmed by future investigations, it can help us locate and understand the physical process behind these rapid variations. 

During our observations, 3C~279 showed an unusual inner jet direction, inconsistent with the prediction of jet precession models by \citet{2011RAA....11...43Q,2012RAA....12...46Q}. The new jet direction, along with the swing (i.e., motion of component A0 along a non-radial trajectory) of the inner jet seen with 43\,GHz data~\citep{Jorstad13}, reflects the non-ballistic nature of the jet and raise the possibility that this phenomenon is associated with a process erratic in nature, similar to the jet of OJ 287~\citep{2012ApJ...747...63A}. Our results indicate that the core brightness temperature at $\lambda$ 1.3 mm is significantly lower than at longer wavelengths \citep[e.g.,][]{2000ApJ...537...91P}. With better {\it uv} coverage and sensitivity in the very near future, it is obvious that VLBI at 1.3 mm will be able to directly image the sub-parsec scale emission. 

\acknowledgments
 High-frequency VLBI work at MIT Haystack Observatory is supported by
grants from the National Science Foundation (NSF). The Submillimeter Array is a joint project between the Smithsonian Astrophysical Observatory and the Academia Sinica Institute of Astronomy and Astrophysics and is funded by the Smithsonian Institution and the Academia Sinica. High-frequency VLBI work at the Arizona Radio Observatory is partially supported through the NSF ATI (AST-0905844) and URO (AST-1140030) programs. This study makes use of 43 GHz VLBA data from the Boston University gamma-ray blazar monitoring program (http://www.bu.edu/blazars/VLBAproject.html), funded by NASA through the Fermi Guest Investigator Program. The National Radio Astronomy Observatory is a facility of the National Science Foundation operated under cooperative agreement by Associated Universities, Inc. L.L. acknowledges the financial support of CONACyT, Mexico (project 104497) DGAPA, UNAM (project IN118110) and the John Simon Guggenheim Memorial Foundation (Fellowship 2010-2011)

{\it Facilities:} \facility{ARO SMT}, \facility{CARMA}, \facility{CSO},
 \facility{JCMT}, \facility{SMA}, \facility{VLBA}

\begin{longtable}[]{ccccccccc}
\caption[Gain-corrected Visibility Amplitudes of 3C 279]{Gain-corrected Visibility Amplitudes of 3C 279}
\label{tab:flux_online} \\
\hline
\multicolumn{1}{c}{Day} & \multicolumn{1}{c}{hh} & \multicolumn{1}{c}{mm} & \multicolumn{1}{c}{Baseline} & \multicolumn{1}{c}{\it u}& \multicolumn{1}{c}{\it v} & \multicolumn{1}{c}{Flux Density} & \multicolumn{1}{c}{$\sigma$} & \multicolumn{1}{c}{Band}\\
\multicolumn{1}{c}{}&\multicolumn{1}{c}{}&  \multicolumn{1}{c}{}&\multicolumn{1}{c}{}&\multicolumn{1}{c}{(M$\lambda$)}&\multicolumn{1}{c}{(M$\lambda$)}&\multicolumn{1}{c}{(Jy)}&\multicolumn
{1}{c}{(Jy)}&\multicolumn{1}{c}{}\\
\hline
\endfirsthead
\multicolumn{4}{l}%
{{\bfseries \tablename\ \thetable{} -- continued}}\\
\hline
\multicolumn{1}{c}{Day} & \multicolumn{1}{c}{hh} & \multicolumn{1}{c}{mm}& \multicolumn{1}{c}{Baseline}& \multicolumn{1}{c}{\it u}& \multicolumn{1}{c}{\it v} &\multicolumn{1}{c}{Flux Density} & \multicolumn{1}{c}{$\sigma$} &\multicolumn{1}{c}{Band} \\
\multicolumn{1}{c}{}&\multicolumn{1}{c}{}&  \multicolumn{1}{c}{}&\multicolumn{1}{c}{}&\multicolumn{1}{c}{(M$\lambda$)}&\multicolumn{1}{c}{(M$\lambda$)}&\multicolumn{1}{c}{(Jy)}&\multicolumn
{1}{c}{(Jy)}&\multicolumn{1}{c}{}\\
\hline
\endhead
\hline
\endfoot
\hline
\endlastfoot
88 & 07 & 26 & CD &     -0.006&     -0.073 &  13.83 &   0.18&high\\  
   & 07 & 26 & SC &   -526.942&    283.963 &   6.14 &   0.12&high\\
   & 07 & 26 & SD &   -526.948&    283.889 &   6.14 &   0.11&high\\
   & 08 & 34 & CD &      0.011&     -0.074 &  13.83 &   0.20&high\\
   & 08 & 34 & CJ &  -2794.721&  -1298.854 &   1.17 &   0.15&high\\
   & 08 & 34 & CP &  -2794.731&  -1298.751 &   1.12 &   0.14&high\\
   & 08 & 34 & DJ &  -2794.732&  -1298.780 &   1.01 &   0.14&high\\
   & 08 & 34 & DP &  -2794.742&  -1298.678 &   1.06 &   0.12&high\\
\end{longtable}
\begin{longtable}{ccccccc}
\caption[Closure phases of 3C 279]{Closure phases of 3C 279}
\label{tab:cphase_online} \\
\hline
\multicolumn{1}{c}{Day} & \multicolumn{1}{c}{hh} & \multicolumn{1}{c}{mm}& \multicolumn{1}{c}{Triangle} &\multicolumn{1}{c}{Closure Phase} & \multicolumn{1}{c}{$\sigma$} &\multicolumn{1}{c}{Band} \\
\multicolumn{1}{c}{} & \multicolumn{1}{c}{} & \multicolumn{1}{c}{} & \multicolumn{1}{c}{}& \multicolumn{1}{c}{(Degree)}&\multicolumn{1}{c}{(Degree)}&\multicolumn{1}{c}{}\\
\hline
\endfirsthead
\multicolumn{4}{l}%
{{\bfseries \tablename\ \thetable{} -- continued}}\\
\hline
\multicolumn{1}{c}{Day} & \multicolumn{1}{c}{hh} & \multicolumn{1}{c}{mm}& \multicolumn{1}{c}{Triangle} &\multicolumn{1}{c}{Closure Phase} & \multicolumn{1}{c}{$\sigma$} &\multicolumn{1}{c}{Band} \\
\multicolumn{1}{c}{} & \multicolumn{1}{c}{} & \multicolumn{1}{c}{} & \multicolumn{1}{c}{}& \multicolumn{1}{c}{(Degree)}&\multicolumn{1}{c}{(Degree)}&\multicolumn{1}{c}{}\\
\hline
\endhead
\hline 
\endfoot
\hline
\endlastfoot
88 & 08 & 34 & CDP &   -11.9 &   5.8 &  high\\
   & 09 & 32 & CDP &    -1.8 &   4.1 &  high\\
   & 08 & 34 & CDJ &    -0.3 &   5.6 &  high\\
   & 09 & 32 & CDJ &    -2.5 &   6.0 &  high\\
   & 10 & 01 & CDJ &     1.2 &   4.0 &  high\\
   & 08 & 34 & CJP &    -3.2 &   7.2 &  high\\
   & 09 & 32 & CJP &     2.7 &   4.7 &  high\\
   & 08 & 34 & DJP &     3.7 &   4.5 &  high\\
\end{longtable}


\begin{thebibliography}{}
\bibitem[Abraham \& Carrara(1998)]{1998ApJ...496..172A} Abraham, Z., \& Carrara, E.~A.\ 1998, \apj, 496, 172 
\bibitem[Agudo et al.(2012)]{2012ApJ...747...63A} Agudo, I., Marscher, 
A.~P., Jorstad, S.~G., et al.\ 2012, \apj, 747, 63
\bibitem[Attridge(2001)]{2001ApJ...553L..31A} Attridge, J.~M.\ 2001, \apjl, 553, L31 
\bibitem[Bloom et al.(2013)]{2013AJ....145...12B} Bloom, S.~D., Fromm, C.~M., \& Ros, E.\ 2013, \aj, 145, 12 
\bibitem[Chatterjee et al.(2008)]{2008ApJ...689...79C} Chatterjee, R., 
Jorstad, S.~G., Marscher, A.~P., et al.\ 2008, \apj, 689, 79
\bibitem[Cheung(2002)]{2002ApJ...581L..15C} Cheung, C.~C.\ 2002, \apjl, 
581, L15
\bibitem[Cohen et al.(1971)]{1971ApJ...170..207C} Cohen, M.~H., Cannon, W., 
Purcell, G.~H., et al.\ 1971, \apj, 170, 207
\bibitem[Collmar et al.(2010)]{2010A&A...522A..66C} Collmar, W., B{\"o}ttcher, M., Krichbaum, T.~P., et al.\ 2010, \aap, 522, A66 
\bibitem[Cotton et al.(1979)]{1979ApJ...229L.115C} Cotton, W.~D., 
Counselman, C.~C., III, Geller, R.~B., et al.\ 1979, \apjl, 229, L115
\bibitem[de Pater \& Perley(1983)]{1983ApJ...273...64D} de Pater, I., \& Perley,
 R.~A.\ 1983, \apj, 273, 64
 
 \bibitem[Doeleman et al.(2009)]{2009ApJ...695...59D} Doeleman, S.~S., Fish, 
V.~L., Broderick, A.~E., Loeb, A., \& Rogers, A.~E.~E.\ 2009, \apj, 695, 59 
\bibitem[Doeleman et al.(2012)]{2012Sci...338..355D} Doeleman, S.~S., Fish, 
V.~L., Schenck, D.~E., et al.\ 2012, Science, 338, 355 
\bibitem[Doeleman et al.(2001)]{2001AJ....121.2610D} Doeleman, S.~S., Shen, Z.-Q., Rogers, A.~E.~E., et al.\ 2001, \aj, 121, 2610 
 \bibitem[Doeleman et al.(2008)]{Doeleman2008} Doeleman, S.~S., Weintroub, J., Rogers, A. E. E., et
  al.\ 2008, \nat, 455, 78
 \bibitem[Fish et al.(2009)]{2009ApJ...692L..14F} Fish, V.~L., Broderick, 
A.~E., Doeleman, S.~S., \& Loeb, A.\ 2009, \apjl, 692, L14
 \bibitem[Fish et al.(2011)]{2011ApJ...727L..36F} Fish, V.~L., Doeleman, S.~S., Beaudoin, C., et al.\ 2011, \apjl, 727, L36
\bibitem[Greve et al.(2002)]{2002A&A...390L..19G} Greve, A., K{\"o}n{\"o}nen, P., Graham, D.~A., et al.\ 2002, \aap, 390, L19
\bibitem[Gurwell et al.(2007)]{2007ASPC..375..234G} Gurwell, M.~A., Peck, A.~B., Hostler, S.~R., Darrah, M.~R., 
\& Katz, C.~A.\ 2007, From Z-Machines to ALMA: (Sub)Millimeter Spectroscopy of Galaxies, 375, 234 
\bibitem[Hayashida et al.(2012)]{2012ApJ...754..114H} Hayashida, M., Madejski, G.~M., Nalewajko, K., et al.\ 2012, \apj, 754, 114 
\bibitem[Homan et al.(2003)]{2003ApJ...589L...9H} Homan, D.~C., Lister, M.~L., Kellermann, K.~I., et al.\ 2003, \apjl, 589, L9 
\bibitem[Homan et al.(2009)]{2009ApJ...696..328H} Homan, D.~C., Lister, M.~L., Aller, H.~D., Aller, M.~F., 
\& Wardle, J.~F.~C.\ 2009, \apj, 696, 328 
\bibitem[Jorstad et al.(2001)]{2001ApJS..134..181J} Jorstad, S.~G., Marscher, A.~P., Mattox, J.~R., et al.\ 2001, \apjs, 134, 181
\bibitem[Jorstad et al.(2004)]{2004AJ....127.3115J} Jorstad, S.~G., Marscher, A.~P., Lister, M.~L., et al.\ 2004, \aj, 127, 3115 
\bibitem[Jorstad et al.(2005)]{2005AJ....130.1418J} Jorstad, S.~G., Marscher, A.~P., Lister, M.~L., et al.\ 2005, \aj, 130, 1418
\bibitem[Jorstad et al. (in prep)]{Jorstad13} Jorstad, S.~G., et al., 2013, in prep. 
\bibitem[Krichbaum et al.(1997)]{1997A&A...323L..17K} Krichbaum, T.~P., Graham, D.~A., Greve, A., et al.\ 1997, \aap, 323, L17
\bibitem[Larionov et al.(2008)]{2008A&A...492..389L} Larionov, V.~M., Jorstad, S.~G., Marscher, A.~P., et al.\ 2008, \aap, 492, 389 
\bibitem[Lee et al.(2008)]{2008AJ....136..159L} Lee, S.-S., Lobanov, A.~P., 
Krichbaum, T.~P., et al.\ 2008, \aj, 136, 159 
\bibitem[Lepp\"{a}nen et al.(1995)]{1995AJ....110.2479L} Lepp\"{a}nen, K.~J., 
Zensus, J.~A., \& Diamond, P.~J.\ 1995, \aj, 110, 2479 
\bibitem[Lu et al.(2011)]{2011MNRAS.418.2260L} Lu, R.-S., Krichbaum, T.~P., 
\& Zensus, J.~A.\ 2011, \mnras, 418, 2260
\bibitem[Lu et al.(2012)]{2012ApJ...757L..14L} Lu, R.-S., Fish, V.~L., Weintroub, J., et al.\ 2012, \apjl, 757, L14 
\bibitem[Marziani et al.(1996)]{1996ApJS..104...37M} Marziani, P., Sulentic, J.~W., Dultzin-Hacyan, D., Calvani, M., 
\& Moles, M.\ 1996, \apjs, 104, 37
\bibitem[Pauliny-Toth et al.(1981)]{1981AJ.....86..371P} Pauliny-Toth, I.~I.~K., Preuss, E., Witzel, A., et al.\ 1981, \aj, 86, 371 
\bibitem[Pearson(1995)]{1995ASPC...82..268P} Pearson, T.~J.\ 1995, in ASP Conf. Ser. 82, Very 
Long Baseline Interferometry and the VLBA, ed., J. A. Zensus, P. J. Diamond, \& P. J. Napier (San Francisco, CA: ASP), 268
\bibitem[Piner et al.(2000)]{2000ApJ...537...91P} Piner, B.~G., Edwards, 
P.~G., Wehrle, A.~E., et al.\ 2000, \apj, 537, 91 
\bibitem[Pi{\'e}tu et al.(2011)]{2011A&A...531L...2P} Pi{\'e}tu, V., di Folco, E., Guilloteau, S., Gueth, F., \& Cox, P.\ 2011, \aap, 531, L2
\bibitem[Qian(2011)]{2011RAA....11...43Q} Qian, S.-J.\ 2011, Research in 
Astronomy and Astrophysics, 11, 43 
\bibitem[Qian(2012)]{2012RAA....12...46Q} Qian, S.-J.\ 2012, Research in 
Astronomy and Astrophysics, 12, 46 
\bibitem[Rogers et al.(1995)]{1995AJ....109.1391R} Rogers, A.~E.~E., 
Doeleman, S.~S., \& Moran, J.~M.\ 1995, \aj, 109, 1391
\bibitem[Taylor(2000)]{2000ApJ...533...95T} Taylor, G.~B.\ 2000, \apj, 533, 95 
\bibitem[Unwin et al.(1989)]{1989ApJ...340..117U} Unwin, S.~C., Cohen, M.~H., Hodges, M.~W., Zensus, J.~A., 
\& Biretta, J.~A.\ 1989, \apj, 340, 117 
\bibitem[Wardle et al.(1998)]{1998Natur.395..457W} Wardle, J.~F.~C., Homan, 
D.~C., Ojha, R., \& Roberts, D.~H.\ 1998, \nat, 395, 457 
\bibitem[Wehrle et al.(2001)]{2001ApJS..133..297W} Wehrle, A.~E., Piner, 
B.~G., Unwin, S.~C., et al.\ 2001, \apjs, 133, 297
\bibitem[Whitney et al.(1971)]{1971Sci...173..225W} Whitney, A.~R., 
Shapiro, I.~I., Rogers, A.~E.~E., et al.\ 1971, Science, 173, 225 
\bibitem[Zavala \& Taylor(2001)]{2001ApJ...550L.147Z} Zavala, R.~T., \& Taylor, G.~B.\ 2001, \apjl, 550, L147 
\end{thebibliography}
\end{document}